\documentclass[conference]{IEEEtran}
\IEEEoverridecommandlockouts

\usepackage{cite}
\usepackage{flushend}
\usepackage{amsmath,amssymb,amsfonts}
\usepackage{graphicx}
\usepackage{booktabs}
\usepackage[ruled,vlined,linesnumbered]{algorithm2e}
\usepackage{array}
\usepackage{siunitx}
\usepackage{subcaption} 
\usepackage{tabularx}
\SetAlgoSkip{smallskip}
\def\BibTeX{{\rm B\kern-.05em{\sc i\kern-.025em b}\kern-.08em
    T\kern-.1667em\lower.7ex\hbox{E}\kern-.125emX}}

\begin{document}

\title{PARGAR: Pilot-Aware Graph Attention for Sparse-Pilot Neural OFDM Receivers}

\author{%
    Sajedeh Norouzi\textsuperscript{1}, Mostafa Rahmani Ghourtani\textsuperscript{2}, Eric Samikwa\textsuperscript{1}, Torsten Braun\textsuperscript{1}\\
    \textsuperscript{1}Institute of Computer Science, University of Bern, Bern, Switzerland.\\
    \textsuperscript{2}5G/6GIC, Institute for Communication Systems (ICS) University of Surrey, U.K. 
    \thanks{
    Corresponding author: Sajedeh Norouzi (Email: sajedeh.norouzi@unibe.ch.) 
    The research was sponsored by the Swiss National Science Foundation (SNSF) Grant No. 219330.}
}

\maketitle

\begin{abstract}
Reliable channel estimation and symbol detection in 5G New Radio (NR), as well as emerging 6G radio systems, rely on demodulation reference signals (DMRS).
The density and placement of these pilots on the orthogonal frequency-division multiplexing (OFDM) time--frequency resource grid affect both receiver reliability and spectral efficiency. Sparse DMRS layouts increase the time--frequency distance between pilot observations and data resource elements (REs), which can make pilot information harder to exploit with purely local processing. This paper proposes \textbf{PARGAR (Pilot-Aware Graph Attention Receiver)}, a graph-attention neural receiver that represents the OFDM resource grid as a graph for bit detection. The graph combines local time--frequency edges with pilot-anchor links from data REs to selected pilot REs, and we further study a channel-scaled pilot-anchor rule based on approximate coherence time and coherence bandwidth. Using Sionna-generated OFDM simulations, we evaluate \textsc{PARGAR} against DeepRx, a convolutional neural network (CNN)-based receiver, across dense, two-symbol sparse, and single-symbol sparse DMRS layouts under channel-mismatch scenarios. Under the considered SISO uncoded setup, channel-scaled \textsc{PARGAR} explicitly injects pilot-aware inductive bias and yields lower BER in selected sparse-pilot scenarios while remaining competitive under dense pilots.
\end{abstract}

\section{Introduction}
\label{sec:introduction}
Fifth-generation (5G) New Radio (NR) and emerging sixth-generation (6G) radio systems rely on demodulation reference signals (DMRS) for channel estimation and coherent detection.
The 5G-Advanced evolution and Release-20 6G study items target mobility, non-terrestrial networks, sensing, and AI-native air interfaces~\cite{11186253}. DMRS density directly controls a practical tradeoff. More pilot resource elements (REs) provide more frequent channel observations and can improve receiver reliability, but they also reduce the number of REs available for data transmission. Reducing pilot density improves the data-RE fraction but increases the time-frequency distance between pilot observations and data REs. In this setting, the receiver must infer channel information over longer pilot-to-data separations, and interpolation errors can affect subsequent equalization and bit detection under delay spread, Doppler, and frequency-selective fading.

Neural-network-based receivers, hereafter referred to as neural receivers, offer an alternative by learning channel estimation and data detection directly from received OFDM resource grids. 
DeepRx showed that CNNs can perform direct bit detection from resource-grid inputs~\cite{honkala2021deeprx}, and subsequent work extended neural receivers toward multi-user multiple-input multiple-output (MIMO) systems, adaptation, continual learning, lightweight implementations, and neuromorphic architectures~\cite{wiesmayr2025standard,obeed2026continual,gupta2026neuromorphic}. These receivers process the grid through convolutional operations, which are effective for local time--frequency neighborhoods. However, sparse pilots create nonlocal pilot-to-data dependencies: a data RE may need information from pilot REs several OFDM symbols or subcarriers away. Capturing such dependencies with only local convolutions can require many layers or large receptive fields.

Graph neural networks (GNNs) provide a complementary inductive bias by propagating information over explicitly defined relationships~\cite{rahmani2026survey}, and graph attention networks (GATs) learn adaptive weights for neighboring nodes~\cite{velivckovic2017graph}. In wireless systems, many GNN methods model users, antennas, links, or resource-allocation relationships for MIMO detection, beamforming, tracking, and scheduling~\cite{yang2020tracking,cammerer2023neural,norouzi2025grinet}. 
Our earlier work used OFDM resource-grid graphs for NR channel estimation~\cite{norouzi2025graphnet}, without performing end-to-end bit detection or introducing pilot-anchor connections between data and pilot REs.
This paper instead uses the OFDM resource grid itself as the graph and explicitly adds pilot-anchor edges to support bit detection.

Robustness under channel mismatch is a second concern. In this paper, channel mismatch refers to test conditions in which the channel profile, terminal speed, or delay spread differs from the training configuration. Neural receivers can be sensitive to the channel distribution used during training~\cite{berke2026site}, and VERITAS~\cite{11282490veritas} showed that such mismatches can degrade receiver performance, while also suggesting that training on harder conditions may generalize better to easier ones. 
These observations lead to three research questions: 1) How to explicitly propagate sparse pilot information to data REs for neural OFDM detection? 2) How to select pilot anchors using channel correlation rather than only grid geometry? 3) How does the resulting graph receiver compare with neural and classical baselines under pilot sparsity and channel mismatch? They are addressed as follows. We propose PARGAR (Pilot-Aware Graph Attention Receiver, a graph-attention receiver for OFDM bit detection in 5G NR- and 6G-oriented radio links, and make the following contributions:
\begin{itemize}
    \item  We introduce a pilot-aware graph-attention neural receiver whose graph combines local time--frequency edges with pilot-anchor edges between data and pilot REs.
    \item We study a channel-scaled anchor-selection rule that normalizes time and frequency separations by approximate coherence time and coherence bandwidth.
    \item We combine practical model-based node features with iterative channel-feature refinement and evaluate \textsc{PARGAR} against DeepRx, least-squares minimum mean-square error (LS-MMSE), and genie-aided MMSE (Genie-MMSE) under dense, two-symbol sparse, and single-symbol sparse pilot layouts with channel-distribution mismatch.
\end{itemize}
\vspace{-1.5 mm}
\section{System Model}
\label{sec:system_model}
\begin{figure}
    \centering
    \includegraphics[width=1\linewidth]{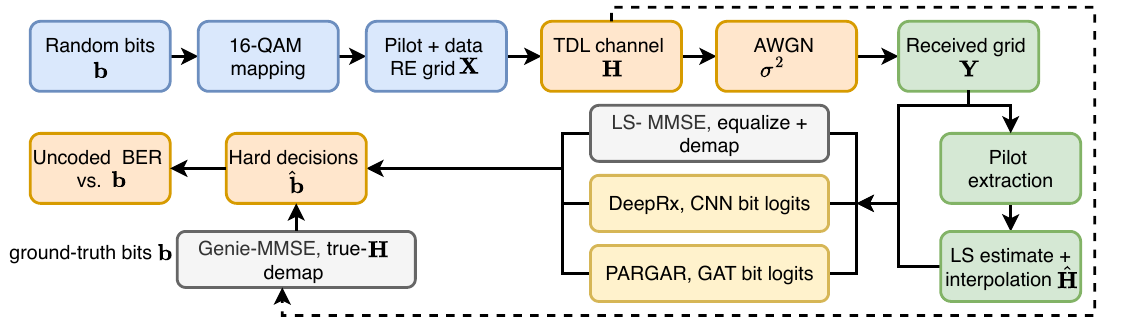}
    \caption{Receiver evaluation pipeline for NR OFDM link.}
    \label{fig:berpipeline}
\end{figure}

\subsection{SISO OFDM Link and Post-FFT Grid}
\label{sec:siso_ofdm_model}
We consider a single-input single-output (SISO) OFDM link with a 5G NR numerology-1 slot as an NR-inspired resource grid. 
As illustrated in Fig.~\ref{fig:pilot_patterns}, the simulated slot contains $N_t=14$ OFDM symbols and $N_f=72=2\times36$ active subcarriers, resulting in $N=N_tN_f=1008$ resource elements (REs).
For a pilot pattern $\mathcal{P}$, let $\Omega_p$ denote the set of pilot REs and $\Omega_d$ denote the complementary data REs. The transmit grid is
\begin{equation}
    x_{t,k} =
    \begin{cases}
    x_p, & (t,k)\in\Omega_p,\\
    s_m, & (t,k)\in\Omega_d,
    \end{cases}
\end{equation}
where $x_p$ is the known unit-power pilot and $s_m$ is a normalized 16-QAM data symbol. Each data RE carries $Q=\log_2(16)=4$ modulation bits.
The propagation channel follows a standardized 3GPP tapped-delay-line (TDL) model with the selected TDL profile, terminal speed, carrier frequency, and delay spread.
Let $a_{\ell}(t)$ and $\tau_{\ell}$ denote the complex gain and delay of path $\ell$. The frequency-domain channel on subcarrier $k$ is computed as
\begin{equation}
    h_{t,k} =
    \sum_{\ell} a_{\ell}(t)
    \exp\!\left(-j2\pi f_k\tau_{\ell}\right),
    \label{eq:freq_channel}
\end{equation}
where $\Delta f$ is subcarrier spacing, and $f_k=k\Delta f$. This yields a single complex channel coefficient for each RE in the OFDM resource grid.
At symbol $t$ and subcarrier $k$, the received signal is
\begin{equation}
    y_{t,k}=h_{t,k}x_{t,k}+n_{t,k},
    \label{eq:rx_model}
\end{equation}
where $h_{t,k}$ is the complex frequency-domain channel, $x_{t,k}$ is either a data symbol or a known pilot symbol, and $n_{t,k}\sim\mathcal{CN}(0,\sigma^2)$ is complex AWGN. Unit average signal power is assumed, and the SNR is swept over
\begin{equation}
    \mathcal{S}=\{0,2,4,\ldots,20\}\ \mathrm{dB},\qquad
    \sigma^2=10^{-\mathrm{SNR}/10}.
\end{equation}
The learning models operate on the noisy post-FFT received resource grid $\mathbf{Y}\in\mathbb{C}^{N_t\times N_f}$, where $[\mathbf{Y}]_{t,k}=y_{t,k}$ denotes the received symbol on the $k$-th subcarrier of the $t$-th OFDM symbol. The term \emph{post-FFT} means that synchronization, cyclic-prefix (CP) removal, and OFDM demodulation by the fast Fourier transform (FFT) are completed before the neural receiver input is formed. The receiver, therefore, observes frequency-domain time--frequency REs rather than time-domain waveform samples. In addition to $\mathbf{Y}$, the receiver is given the pilot locations, transmitted pilot symbols, and noise variance $\sigma^2$, which is used for SNR conditioning and MMSE-based feature construction.
\subsection{Pilot Patterns and Pilot-Aided Channel Initialization}
\label{sec:pilot_patterns}
\begin{figure}
    \centering
    \includegraphics[width=1\linewidth]{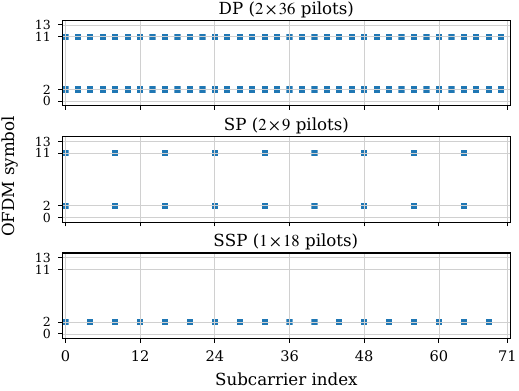}
    \caption{DMRS pilot layouts: DP, SP, and SSP.}
    \label{fig:pilot_patterns}
\end{figure}

Fig.~\ref{fig:pilot_patterns} illustrates the three DMRS-like layouts used to study pilot-density and pilot-placement effects. Pilots are placed at fixed, known REs, and each pilot RE carries the same known unit-power symbol. The dense pilot pattern (DP) places pilots on two OFDM symbols, 2 and 11, with 36 pilot subcarriers per pilot symbol, resulting in $2\times36=72$ pilot REs.
The sparse pilot pattern (SP) retains the same two pilot symbols but uses nine pilot subcarriers per pilot symbol, reducing the pilot count to $2\times9=18$ REs and increasing the available data REs from 936 to 990. We also evaluate a single-symbol sparse pilot (SSP) layout, which places 18 pilots on a single OFDM symbol with pilot subcarriers spaced every 4 subcarriers; this maintains the same pilot count as SP but changes the temporal placement of the pilot observations.
The LS channel estimate on pilot REs is
\begin{equation}
    \hat{h}^{\mathrm{LS}}_{t,k}=\frac{y_{t,k}}{x_p},
    \qquad (t,k)\in \Omega_p.
    \label{eq:ls}
\end{equation}
The pilot estimates are interpolated in frequency and time to obtain a full-grid channel estimate $\hat{h}_{t,k}$ for all REs. This LS interpolation provides a practical receiver-side channel initialization from the available pilots without using true channel state information (CSI). In frequency, the available pilot-subcarrier estimates are interpolated to all $72=2\times36$ active subcarriers on each pilot OFDM symbol. In time, DP and SP use linear interpolation between the two pilot-bearing OFDM symbols and nearest-neighbor extrapolation outside them, while SSP copies the single pilot-symbol estimate across time after frequency interpolation. This same full-grid estimate initializes DeepRx, \textsc{PARGAR}, and the LS-MMSE reference. The LS-MMSE receiver is the practical classical reference and applies single-tap MMSE equalization using the interpolated estimate:
\begin{equation}
    \hat{x}^{\mathrm{MMSE}}_{t,k} =
    \frac{\hat{h}^{*}_{t,k}y_{t,k}}
    {|\hat{h}_{t,k}|^2+\sigma^2+\epsilon},
    \label{eq:mmse_eq}
\end{equation}
where $\epsilon$ is a small numerical constant. Genie-MMSE uses the true simulated channel only as a perfect-CSI linear MMSE reference and is not an implementable receiver.
We assume ideal synchronization and carrier-frequency-offset compensation before the post-FFT grid is formed. Channel coding is outside the scope of this study; the reported uncoded BER therefore isolates receiver-side channel estimation and bit-detection behavior for the modulation bits. Fig.~\ref{fig:berpipeline} summarizes the receiver evaluation pipeline used in this work.
\section{Pilot-Aware Graph Attention Receiver}
\label{sec:pargar}
\begin{figure}
    \centering
    \includegraphics[width=1\linewidth]{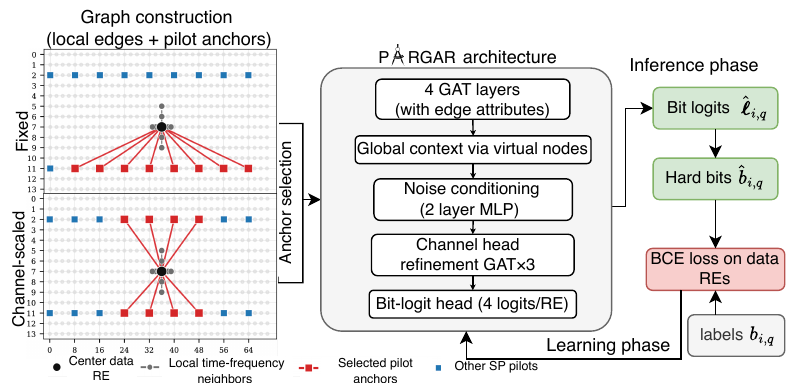}
    \caption{\textsc{PARGAR} graph-attention receiver architecture.}
    \label{fig:pargararchitecture}
\end{figure}

As shown in Fig.~\ref{fig:pargararchitecture}, \textsc{PARGAR} first constructs LS-interpolated and MMSE-derived features from the received grid, then maps the full OFDM slot to a graph whose nodes are REs and whose edges encode local time--frequency relations and pilot-anchor links. This representation is motivated by the sparse-pilot setting: local edges capture the neighboring OFDM structure, while pilot-anchor edges provide data REs with direct message-passing paths to selected pilot REs. These direct links reduce the number of graph-attention steps required for pilot information to reach distant data REs, and are expected to be especially useful when pilot density is low. Graph attention layers propagate pilot information across this graph, iterative refinement updates data-node channel features, and 
a final MLP produces bit logits, i.e., raw bit scores before sigmoid activation.
The receiver uses a pattern-aware data mask and a 9-dimensional node-feature vector.
For each RE, the graph input feature is
\begin{equation}
\begin{aligned}
    \mathbf{r}_{t,k} = \big[
    &\Re\{y_{t,k}\},\,
    \Im\{y_{t,k}\},\,
    \Re\{\hat{h}_{t,k}\},\,
    \Im\{\hat{h}_{t,k}\},\\
    &\Re\{\hat{x}^{\mathrm{MMSE}}_{t,k}\},\,
    \Im\{\hat{x}^{\mathrm{MMSE}}_{t,k}\},\,
    |\hat{h}_{t,k}|,\,
    \log\sigma^2,\,
    \mathbb{1}_{(t,k)\in\Omega_p}
    \big],
\end{aligned}
\label{eq:mmse_node_features}
\end{equation}
The first four entries provide the received symbol and the LS-interpolated channel estimate. The MMSE-derived entries provide a coarse model-based symbol estimate computed from the practical single-tap equalizer in \eqref{eq:mmse_eq}, using the LS-interpolated channel estimate rather than the true channel. The magnitude $|\hat{h}_{t,k}|$ supplies channel-strength information, $\log\sigma^2$ provides a noise-level feature, and the last entry, $\mathbb{1}_{(t,k)\in\Omega_p}$, is a binary pilot indicator that identifies pilot REs.
The graph contains two connectivity classes. First, local time--frequency edges connect each node to nearby REs using frequency offsets $\{1,2,3\}$ and time offsets $\{1,2\}$, together with a self-edge, which connects a node to itself and preserves its own feature information during message passing. A node at RE $(t,k)$ is therefore connected to same-symbol neighbors $k\pm1$, $k\pm2$, and $k\pm3$, and to same-subcarrier neighbors $t\pm1$ and $t\pm2$, whenever the corresponding neighboring REs exist. These edges capture short-range time--frequency correlation in the post-FFT grid. Second, pilot-anchor edges connect every data node to selected pilot nodes. Both local and pilot-anchor connections are bidirectional, allowing pilot information to flow to data nodes and data-node embeddings to attend back to selected pilot anchors. The number of pilot-anchor edges grows with the number of data REs and the anchor count $A_p$.
Each edge carries a six-dimensional edge feature
\begin{equation}
    \mathbf{e}_{i,j}=
    [\Delta t_{i,j},\Delta f_{i,j},\mathbf{o}_{i,j}],
    \qquad
    \mathbf{o}_{i,j}=\psi(\tau_{i,j})\in\{0,1\}^{4},
\end{equation}
where $\Delta t_{i,j}$ and $\Delta f_{i,j}$ are the normalized time and
frequency displacements from node $i$ to node $j$. 
Here, $\tau_{i,j}\in\{\mathrm{self},\mathrm{frequency},\mathrm{time},\mathrm{pilot}\}$ denotes the edge category, and $\psi(\cdot)$ maps this category to the corresponding four-dimensional categorical basis vector. Thus, $\mathbf{o}_{i,j}$ has one nonzero entry indicating whether the edge is a self, frequency, time, or pilot-anchor connection.
For example, a pilot-anchor edge is encoded as
$\mathbf{o}_{i,j}=[0,0,0,1]$. 
\begin{table}[t]
\caption{Receiver architectures and training parameters.}
\label{tab:receiver_params}
\scriptsize
\centering
\begin{tabularx}{\columnwidth}{
>{\raggedright\arraybackslash}p{0.42\columnwidth}
X}
\toprule
\textbf{Parameter} & \textbf{Value} \\
\midrule
    \multicolumn{2}{l}{\textbf{PARGAR}}\\
    Node features & 9 (Rx, LS, MMSE, $|\hat{h}|$, $\log\sigma^2$, pilot flag) \\
    Hidden dimension & 256 \\
    Attention heads/ Initial GAT layers & 4 \\
    Refinement iterations & 3 \\
    Virtual nodes & 8 \\
    Graph connectivity &
    Frequency offsets $\{1,2,3\}$;\\
    & Time offsets $\{1,2\}$; pilot anchors \\
    Anchor count ($A_p$) & $\{4,8\}$ \\
    Trainable parameters & 1.09M \\
    Number of edges &
    DP-A4/A8: 16\,840 / 24\,328;\\
    & SP-A4/A8: 18\,060 / 25\,980;\\
    & SSP-A4/A8: 18\,054 / 25\,974 \\
    \multicolumn{2}{l}{\textbf{DeepRx}}\\
    Architecture & 11 depthwise-separable residual blocks \\
    Input & 4-channel resource grid \\
    Trainable parameters & 1.22M \\
\bottomrule
\end{tabularx}
\end{table}
For a data node $v_i=(t_i,k_i)$ and a pilot node $v_p=(t_p,k_p)$, pilot anchors can be selected using either a fixed geometric rule or a channel-scaled rule. The fixed rule ranks pilots using normalized time--frequency separation,
\begin{equation}
    d_{\mathrm{fix}}(v_i,v_p)=
    \left(\alpha_t\frac{|t_i-t_p|}{N_t-1}\right)^2+
    \left(\frac{|k_i-k_p|}{N_f-1}\right)^2,
    \label{eq:fixed_anchor}
\end{equation}
with $\alpha_t=5$.
This value is used as an empirically chosen fixed-anchor baseline that gives stronger weight to normalized OFDM-symbol separation than to normalized subcarrier separation on the considered OFDM grid. 
The channel-scaled rule instead normalizes the same separations by approximate coherence time and coherence bandwidth, so that pilot selection better reflects the expected channel-correlation structure:
\begin{equation}
    d_{\mathrm{scl}}(v_i,v_p)=
    \left(\frac{|t_i-t_p|\bar{T}}{T_c}\right)^2+
    \left(\frac{|k_i-k_p|\Delta f}{B_c}\right)^2,
    \label{eq:scl_anchor}
\end{equation}
where $\bar{T}=T_{\mathrm{slot}}/N_t$ is the OFDM-symbol duration approximation, $T_{\mathrm{slot}}=\SI{0.5}{ms}$ is the slot duration for NR numerology $\mu=1$ and $N_t=14$ is the number of OFDM symbols per slot. $T_c$ and $B_c$ are approximated as 
\begin{equation}
    T_c \approx \frac{0.423}{f_D},\qquad
    f_D=\frac{v f_c}{c},\qquad
    B_c \approx \frac{1}{5\tau_{\mathrm{rms}}}.
    \label{eq:coherence_params}
\end{equation}
Here, $v$ is the terminal speed, $f_c$ is the carrier frequency, $c$ is the speed of light, and $\tau_{\mathrm{rms}}$ is the delay spread. The coefficient 0.423 corresponds to a conventional coherence-time approximation. 
In this study, the channel-scaled graph assumes coarse Doppler and delay-spread side information, meaning approximate values of the terminal speed and RMS delay spread rather than exact test-channel parameters. These values are taken from the training-side configuration to isolate the effect of the graph-construction rule. The reported BER is evaluated only on the held-out test configurations in Table~\ref{tab:experiment_splits}, where the channel profile, speed, or delay spread can differ from the graph-construction parameters. Thus, the rule does not assume exact knowledge of the test-channel parameters. The graph topology is constructed once per experiment using the stated parameters and reused for all corresponding held-out test conditions. The $A_p$ pilots with the smallest $d_{\mathrm{scl}}(v_i,v_p)$ are used as pilot anchors.                         

The receiver uses graph attention layers with a hidden dimension $d_h$ and four attention heads. A first stack of four graph attention network (GAT) layers maps the regular RE-node feature matrix to embeddings $\mathbf{S}\in\mathbb{R}^{N\times d_h}$, where $N=N_tN_f$ is the number of RE nodes. To inject global slot context, \textsc{PARGAR} uses $K=8$ virtual nodes. These nodes are not physical REs; they are learned global tokens that aggregate information from the entire OFDM slot and then broadcast the resulting context back to the regular RE nodes. Let $\mathbf{Q}\in\mathbb{R}^{K\times d_h}$ denote the trainable virtual node query matrix. The rows of $\mathbf{Q}$ are model parameters, initialized randomly and optimized jointly with the rest of the receiver during training. They are shared across all slots and are not derived from pilots or data symbols. For each received slot, these learned queries attend to the regular-node embeddings $\mathbf{S}$:
\begin{align}
    \mathbf{U} &= \mathrm{LayerNorm}\!\left( \mathbf{Q} +
    \mathcal{A}_{\mathrm{agg}}(\mathbf{Q},\mathbf{S},\mathbf{S})
    \right),
    \label{eq:supernode_aggregate}\\
    \mathbf{S}^{\mathrm{glob}} &=
    \mathrm{LayerNorm}\!\left( \mathbf{S}+
    \mathcal{A}_{\mathrm{bcast}}(\mathbf{S},\mathbf{U},\mathbf{U})
    \right).
    \label{eq:supernode_broadcast}
\end{align}
Here, $\mathcal{A}_{\mathrm{agg}}(\mathbf{Q},\mathbf{S},\mathbf{S})$ is a multi-head attention operation in which the virtual node queries attend to all regular RE-node embeddings, producing slot-level summaries $\mathbf{U} \in\mathbb{R}^{K\times d_h}$. The broadcast operation $\mathcal{A}_{\mathrm{bcast}}(\mathbf{S},\mathbf{U},\mathbf{U})$ then lets each regular RE node attend to these summaries, producing $\mathbf{S}^{\mathrm{glob}}\in\mathbb{R}^{N\times d_h}$. The $i$th row $\mathbf{s}^{\mathrm{glob}}_i$ is the globally conditioned embedding of RE node $i$.
\begin{equation}
    \bar{\mathbf{s}}_i
    =
    \mathbf{s}^{\mathrm{glob}}_i
    +
    g_{\phi}(\log\sigma^2),
    \label{eq:snr_conditioning}
\end{equation}
where $g_{\phi}(\cdot)$ is a learnable two-layer MLP that maps the log-noise variance to a $d_h$-dimensional embedding. This conditioning gives the detector an explicit representation of the operating noise level.

The model then performs iterative refinement. At each refinement step, a soft channel head predicts $[\Re\{\tilde{h}_i\},\Im\{\tilde{h}_i\}]$ from the current embedding for each data node. The channel entries of the corresponding data-node features are updated using this soft estimate, while pilot-node features remain anchored by the pilot-derived LS estimates. A refinement GAT stack is then applied to update the node embeddings. After the final refinement step, a two-layer MLP produces four-bit logits at each data node:
\begin{equation}
    \hat{\boldsymbol{\ell}}^{\mathrm{PARGAR}}_i =
    f_{\phi}^{\mathrm{bit}}(\bar{\mathbf{s}}_i),
    \qquad i\in\Omega_d .
\end{equation}
The training objective is the bitwise binary cross-entropy (BCE):
\begin{equation}
    \mathcal{L}_{\mathrm{PARGAR}}
    =
    \frac{1}{4|\Omega_d|}
    \sum_{i\in\Omega_d}
    \sum_{q=1}^{4}
    \mathrm{BCE}
    \left(b_{i,q},\hat{\ell}^{\mathrm{PARGAR}}_{i,q}\right),
\end{equation}
where $b_{i,q}\in\{0,1\}$ denotes the transmitted ground-truth bit for bit position $q$ of data RE $i$. Algorithm~\ref{alg:pargar} summarizes the complete \textsc{PARGAR} pipeline from pilot-aided channel initialization to graph construction, refinement, and bit-logit prediction.
\begin{algorithm}[hbt!]
\caption{\textsc{PARGAR} workflow}
\label{alg:pargar}
\footnotesize
\KwData{Received grid $\mathbf{Y}$, pilot set $\Omega_p$, data set $\Omega_d$, pilot symbol $x_p$, noise variance $\sigma^2$, pilot-anchor count $A_p$, anchor rule $\rho\in\{d_{\mathrm{fix}},d_{\mathrm{scl}}\}$}
\KwResult{Data-node bit logits $\hat{\boldsymbol{\ell}}^{\mathrm{PARGAR}}$}

$\hat{h}^{\mathrm{LS}}_{t,k} \gets y_{t,k}/x_p,\quad (t,k)\in\Omega_p$\;
$\hat{\mathbf{H}} \gets \mathrm{Interpolate}(\hat{\mathbf{H}}^{\mathrm{LS}},\Omega_p)$\;
$\mathcal{V} \gets \{(t,k):0\leq t<N_t,\ 0\leq k<N_f\}$\;
$\mathcal{E} \gets$ self edges and local time--frequency edges\;

\For{$i\in\Omega_d$}{
    Add bidirectional pilot-anchor edges between $i$ and the $A_p$ pilot nodes with smallest distance under $\rho$\;
}

$\mathbf{r}_{t,k} \gets$ node-feature vector from \eqref{eq:mmse_node_features}, using \eqref{eq:mmse_eq}\;
$\mathbf{S} \gets \mathrm{GAT}_{1}(\mathbf{r},\mathcal{E})$\;
$\mathbf{Q}\in\mathbb{R}^{K\times d_h} \gets$ learned receiver parameters for the $K$ virtual node queries\;
$\mathbf{U} \gets \mathrm{LayerNorm}\!\left(\mathbf{Q}+\mathcal{A}_{\mathrm{agg}}(\mathbf{Q},\mathbf{S},\mathbf{S})\right)$\;
$\mathbf{S}^{\mathrm{glob}} \gets \mathrm{LayerNorm}\!\left(\mathbf{S}+\mathcal{A}_{\mathrm{bcast}}(\mathbf{S},\mathbf{U},\mathbf{U})\right)$\;
$\bar{\mathbf{s}}_i \gets \mathbf{s}^{\mathrm{glob}}_i + g_{\phi}(\log\sigma^2),\quad i\in\mathcal{V}$\;

\For{each refinement iteration}{
    $[\Re\{\tilde{h}_i\},\Im\{\tilde{h}_i\}]
    \gets \mathrm{ChannelHead}(\bar{\mathbf{s}}_i),\quad i\in\Omega_d$\;
    Update data-node channel features with $\tilde{h}_i$ and keep pilot-node channel features \;
    Apply the refinement GAT stack to update $\bar{\mathbf{s}}_i$\;
}

$\hat{\boldsymbol{\ell}}^{\mathrm{PARGAR}}_i
\gets f_{\phi}^{\mathrm{bit}}(\bar{\mathbf{s}}_i),
\quad i\in\Omega_d$\;

\Return{$\hat{\boldsymbol{\ell}}^{\mathrm{PARGAR}}$}
\end{algorithm}
\section{Experimental Setup}
The experiments compare the proposed \textsc{PARGAR} receiver with three references: DeepRx, a CNN-based neural receiver baseline; LS-MMSE, a practical classical receiver that uses an interpolated LS channel estimate; and Genie-MMSE, a perfect-CSI linear MMSE reference that uses the true simulated channel. We first describe the DeepRx baseline and then detail the shared simulation setup, training, and test scenarios. 
\label{sec:experimentsetup}
\subsection{DeepRx Baseline}
\label{sec:deeprx}
DeepRx~\cite{honkala2021deeprx} is used as the CNN-based neural receiver baseline. Unlike \textsc{PARGAR}, it processes the OFDM grid through convolutional operations. To keep the comparison controlled, DeepRx receives the same post-FFT received grid and LS-interpolated channel estimate as \textsc{PARGAR}:
\begin{equation}
    \mathbf{z}^{\mathrm{DRX}}_{t,k}
    =
    \left[
    \Re\{y_{t,k}\},
    \Im\{y_{t,k}\},
    \Re\{\hat{h}_{t,k}\},
    \Im\{\hat{h}_{t,k}\}
    \right].
\end{equation}
The network outputs four-bit logits per RE,
\begin{equation}
    \hat{\boldsymbol{\ell}}^{\mathrm{DRX}}_{t,k}
    =
    f_{\theta}^{\mathrm{DRX}}(\mathbf{Y},\hat{\mathbf{H}})_{t,k}
    \in\mathbb{R}^{4}.
\end{equation}
The implementation uses a $3\times3$ convolutional stem, followed by 11 preactivation residual blocks with depthwise separable convolutions, batch normalization, ReLU activations, and temporal and frequency-domain dilations. A final $1\times1$ convolution produces the bit logits. As with \textsc{PARGAR}, only data REs contribute to the training loss:
\begin{equation}
    \mathcal{L}_{\mathrm{DRX}}
    =
    \frac{1}{4|\Omega_d|}
    \sum_{(t,k)\in\Omega_d}
    \sum_{q=1}^{4}
    \mathrm{BCE}
    \left(b_{t,k,q},\hat{\ell}^{\mathrm{DRX}}_{t,k,q}\right).
\end{equation}
Table~\ref{tab:receiver_params} summarizes the receiver architecture parameters. \textsc{PARGAR} has 1.09M trainable parameters and operates on a full-slot graph whose edge count depends on pilot density, pilot placement, and anchor count $A_p$. Increasing $A_p$ adds pilot-anchor edges, and the sparse layouts have more edges than DP because they contain more data REs. DeepRx uses an 11-block depthwise-separable convolutional architecture with 1.22M trainable parameters.
\begin{table}[t]
\caption{Physical-layer, Sionna data-generation, and training parameters.}
\label{tab:sim_params}
\scriptsize
\centering
\begin{tabularx}{\columnwidth}{
>{\raggedright\arraybackslash}p{0.37\columnwidth}
>{\centering\arraybackslash}p{0.14\columnwidth}
X}
\toprule
\textbf{Parameter} & \textbf{Notation} & \textbf{Value} \\
\midrule
    \multicolumn{3}{l}{\textbf{Physical-Layer Configuration}}\\
    Link setting & -- & SISO, post-FFT resource grid \\
    Subcarrier spacing & $\Delta f$ & \SI{30}{kHz} \\
    Slot duration & $T_{\mathrm{slot}}$ & \SI{0.5}{ms} \\
    Grid size & $N_t \times N_f$ & $14 \times 72$ REs  \\
    Modulation order & $M$ & 16-QAM \\
    Bits per RE & $Q$ & 4 \\
    SNR range & $\mathcal{S}$ & 0--20 dB (2 dB step) \\ 
    Pilot symbol & $x_p$ & $(1+j)/\sqrt{2}$ \\
    Dense pilot pattern & DP & 72 pilot REs, 936 data REs \\
    Sparse pilot pattern & SP & 18 pilot REs, 990 data REs \\
    Single-symbol sparse pilot & SSP & 18 pilot REs, 990 data REs \\ \hline
    \multicolumn{3}{l}{\textbf{Sionna Channel and Data Generation}}\\
    Channel model & -- & 3GPP TDL-A/B/C/D/E \\
    Carrier frequency & $f_c$ & \SI{3.5}{GHz} \\
    User speed & $v$ & Scenario dependent \\
    Delay spread & $\tau_{\mathrm{rms}}$ & Scenario dependent \\
    TDL time samples & -- & 14 \\
    TDL path-gain sampling rate & $f_s^{\mathrm{TDL}}$ & \SI{28}{kHz} \\
    Noise model & -- & AWGN, $\sigma^2=10^{-\mathrm{SNR}/10}$ \\
    Monte Carlo slots & -- & 500 per channel condition per SNR \\ \hline

    \multicolumn{3}{l}{\textbf{Neural-Receiver Training and Evaluation}}\\
    Epochs & -- & 80 \\
    Batch size & -- & DeepRx: 32; PARGAR: 16 \\
    Train/validation split & -- & 85\% / 15\%  \\
    Loss function & $\mathcal{L}$ & Binary cross-entropy (BCE) \\
    Evaluation metric & $\mathrm{BER}$ & Uncoded BER on data REs \\ 
\bottomrule
\end{tabularx}
\end{table}
We used Sionna~\cite{hoydis2022sionna} to generate standardized 3GPP TDL channel realizations for the SISO OFDM link. For each channel condition in Table~\ref{tab:experiment_splits}, Sionna generates TDL path gains over the 14 OFDM-symbol time samples of a slot. The corresponding frequency-domain channel coefficients on the 72 active subcarriers are then computed using \eqref{eq:freq_channel}, yielding the post-FFT channel grid used in \eqref{eq:rx_model}. For each experiment, pilot pattern, channel condition, and SNR point, we generate 500 independent OFDM slots offline. Thus, each BER point is estimated from at least $500\times936\times4=1.872\times10^6$ bits for DP and $500\times990\times4=1.98\times10^6$ bits for SP and SSP, which is sufficient for the smallest BER values reported in this study. Random information bits are mapped to normalized Gray-coded 16-QAM symbols, pilots are inserted according to the selected DMRS-like layout (DP/SP/SSP), and complex AWGN is added according to the target SNR. 
The physical-layer, Sionna data-generation, and training parameters are summarized in Table~\ref{tab:sim_params}.
Pilot observations are converted to LS estimates and linearly interpolated over the full grid. The resulting channel estimate is shared by LS-MMSE, DeepRx, and \textsc{PARGAR}; Genie-MMSE uses the true simulated channel only as a perfect-CSI linear MMSE reference.
For each experiment, neural receivers are trained only on the training configurations in Table~\ref{tab:experiment_splits}. The generated training slots are split into 85\% for training and 15\% for validation, and the checkpoint with the lowest validation BCE is selected for testing. Test configurations are never used for training or checkpoint selection. All reported uncoded BER results are obtained on the held-out test configurations in Table~\ref{tab:experiment_splits}.
The experiments evaluate DeepRx, \textsc{PARGAR}, LS-MMSE, and Genie-MMSE using uncoded BER on data REs:
\begin{equation}
    \mathrm{BER}
    =
    \frac{1}{4|\Omega_d|}
    \sum_{i\in\Omega_d}\sum_{q=1}^{4}
    \mathbb{1}\{\hat{b}_{i,q}\ne b_{i,q}\},
    \label{eq:ber}
\end{equation}
where $\mathbb{1}\{\cdot\}$ is the indicator function, equal to one if the detected bit differs from the transmitted bit and zero otherwise.
For reference, the pilot overhead can be related to an uncoded effective spectral-efficiency proxy:
\begin{equation}
    R_{\mathrm{eff}} = \frac{|\Omega_d|}{N_t \times N_f} \log_2(M)
    \left(1-\mathrm{BER}\right), \label{eq:effective_se}
\end{equation}
where $|\Omega_d|/N$ is the data-RE fraction after pilot insertion, and $M$ is the modulation order. This quantity is not used as a primary evaluation metric in this work, but it illustrates how reducing pilot density can increase the available data-RE fraction while receiver errors reduce the effective uncoded throughput.

\begin{table*}[t]
\centering
\caption{Training and test splits for the three experiments.}
\label{tab:experiment_splits}
\begin{tabular}{cll}
    \toprule
    Experiment & Training configuration & Test configuration \\ \midrule
    Exp1 & TDL-A, speed \SI{10}{m/s}, delay spread \SI{400}{ns} & TDL-A/B/C/D/E, speed \SI{10}{m/s}, delay spread \SI{400}{ns} \\
    Exp2 & TDL-D, speeds \SIlist{18;19;20}{m/s}, delay spread \SI{400}{ns} & TDL-D, speeds \SIlist{1;16;17}{m/s}, delay spread \SI{400}{ns} \\
    Exp3 & TDL-B, speed \SI{2}{m/s}, delay spreads \SIlist{400;450;500}{ns} & TDL-B, speed \SI{2}{m/s}, delay spreads \SIlist{10;50;80}{ns} \\
\bottomrule
\end{tabular}
\end{table*}

The simulation setup follows the VERITAS evaluation framework~\cite{11282490veritas}. Three experiments evaluate robustness under changes in channel profile, user mobility, and delay spread. Table~\ref{tab:experiment_splits} summarizes the dataset splits. Exp1 trains on TDL-A and evaluates across TDL-A/B/C/D/E, so the test set includes the training profile as well as additional profiles. Exp2 evaluates speeds different from the training speeds while keeping the TDL profile and delay spread fixed. Exp3 evaluates delay-spread values different from the training values while keeping the TDL profile and speed fixed.

The primary pilot-placement study considers all combinations of the three experiments, three pilot patterns (DP/SP/SSP), and two pilot-anchor values ($A_p=4$ and $A_p=8$). DeepRx is trained for each experiment and for each pilot-pattern dataset. For graph-anchor variants that share the same experiment and pilot-pattern dataset, including different anchor counts and anchor rules, the corresponding DeepRx checkpoint is reused. 
For all evaluations, the checkpoint with the best validation performance is used to measure uncoded BER.
To further investigate anchor placement, the channel-scaled metric in \eqref{eq:scl_anchor} is evaluated for the two sparse layouts, SP and SSP, since the DP results indicate that anchor selection is not a limiting factor when pilots are already dense. In the SP BER comparison, each plotted \textsc{PARGAR} anchor rule reports a post-evaluation summary: the lower average BER result between $A_p=4$ and $A_p=8$ for that experiment and anchor rule. This selection is used only to keep the BER plots readable; the separate anchor-ablation figure reports both anchor counts.
\vspace{-2 mm}
\section{Results and Discussion}
\label{sec:results}
This section reports the uncoded BER performance of DeepRx, \textsc{PARGAR}, LS-MMSE, and Genie-MMSE under the three generalization experiments. Results are presented first for the DP layout, then for the two sparse layouts, SP and SSP, since pilot-anchor graph propagation is expected to be more useful when pilot observations are sparse, and data REs are farther from direct pilot measurements. For each experiment, BER curves are averaged over the corresponding held-out test conditions in Table~\ref{tab:experiment_splits}. 
In all BER figures, the horizontal axis shows the SNR in dB and the vertical axis shows the uncoded BER on a logarithmic scale. Each point is computed by evaluating the selected checkpoint on held-out test slots at the corresponding SNR and averaging bit errors over all data REs and test channel conditions. Lower curves indicate better detection performance, and the separation between curves shows the relative robustness of each receiver under the corresponding generalization setting.
When reporting average relative BER differences, we use the ratio between the average BER difference and the DeepRx average BER over the same SNR--test-condition set.
\begin{figure*}[t]
    \centering
    \begin{subfigure}[t]{0.32\textwidth}
        \centering
        \includegraphics[width=\linewidth]{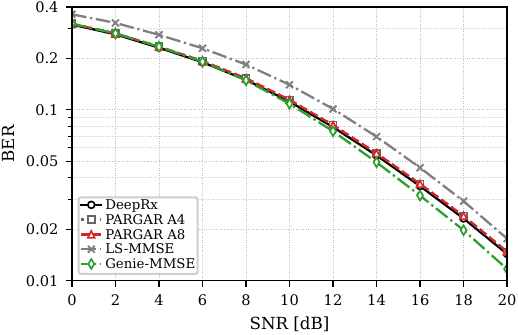}
        \caption{Exp1}
        \label{fig:exp1_p0_ber}
    \end{subfigure}
    \hfill
    \begin{subfigure}[t]{0.32\textwidth}
        \centering
        \includegraphics[width=\linewidth]{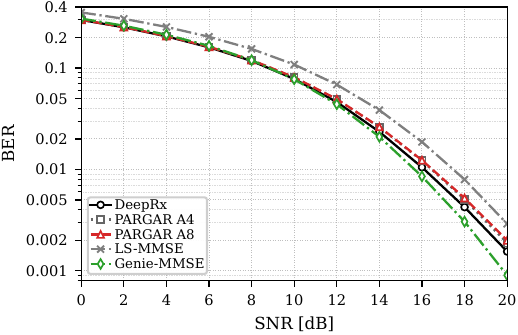}
        \caption{Exp2}
        \label{fig:exp2_p0_ber}
    \end{subfigure}
    \hfill
    \begin{subfigure}[t]{0.32\textwidth}
        \centering
        \includegraphics[width=\linewidth]{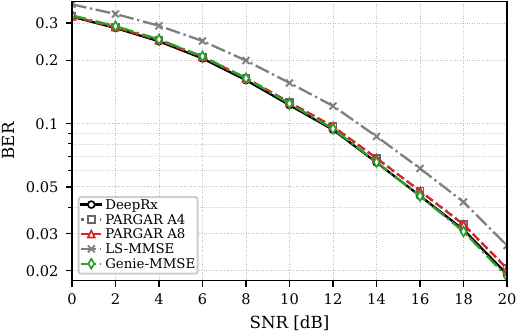}
        \caption{Exp3}
        \label{fig:exp3_p0_ber}
    \end{subfigure}

    \caption{Dense-pilot BER under channel-profile, mobility, and delay-spread mismatch for Exp1-Exp3.}
    \label{fig:dp_ber_comparison}
\end{figure*}

Fig.~\ref{fig:dp_ber_comparison} reports the BER curves for the dense pilot layout. In Fig.~\ref{fig:dp_ber_comparison}(a), corresponding to Exp1, DeepRx and \textsc{PARGAR} remain closely matched across the evaluated SNR range. The best \textsc{PARGAR} configuration differs from DeepRx by only 1.66\% on average, indicating that, under the Exp1 channel-profile evaluation, both receivers can effectively exploit the frequent pilot observations. Fig.~\ref{fig:dp_ber_comparison}(b) shows the Exp2 speed-mismatch case. The learned receivers again produce similar BER curves, with an average relative difference of 1.54\%. This behavior is consistent with dense pilots providing sufficient temporal channel observations for both CNN-based processing and pilot-aware graph-based message passing to track the mobility changes under consideration.
Fig.~\ref{fig:dp_ber_comparison}(c) presents the Exp3 delay-spread mismatch results. The gap between DeepRx and \textsc{PARGAR} remains small, with an average relative BER difference of 1.89\%. Thus, even when the test delay spread differs from the training configuration, dense pilots provide enough frequency-domain channel information for stable interpolation and competitive learned detection. Overall, the dense-pilot results suggest that explicit pilot-anchor connectivity provides limited additional benefit when pilot observations are already frequent. LS-MMSE serves as the practical classical reference, while Genie-MMSE provides a perfect-CSI linear equalization reference.
\begin{figure*}[t]
    \centering
    \begin{subfigure}[t]{0.32\textwidth}
        \centering
        \includegraphics[width=\linewidth]{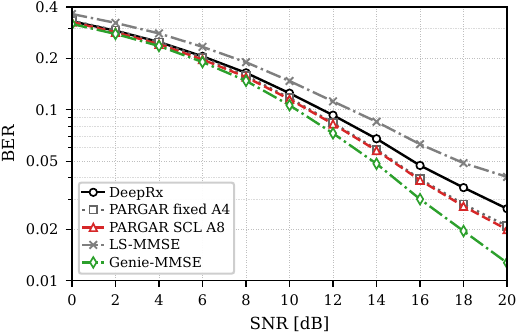}
        \caption{Exp1}
        \label{fig:exp1_p2_ber}
    \end{subfigure}
    \hfill
    \begin{subfigure}[t]{0.32\textwidth}
        \centering
        \includegraphics[width=\linewidth]{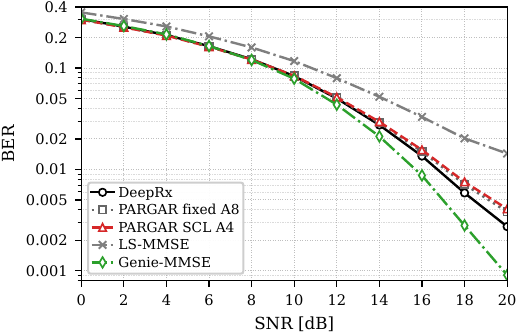}
        \caption{Exp2}
        \label{fig:exp2_p2_ber}
    \end{subfigure}
    \hfill
    \begin{subfigure}[t]{0.32\textwidth}
        \centering
        \includegraphics[width=\linewidth]{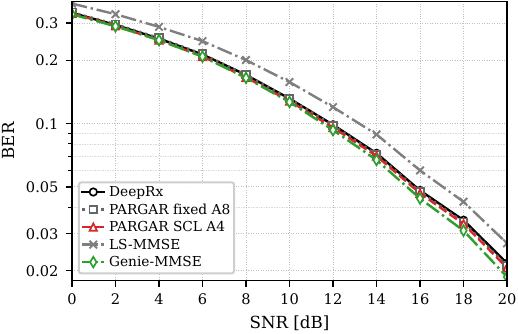}
        \caption{Exp3}
        \label{fig:exp3_p2_ber}
    \end{subfigure}
    \caption{Sparse-pilot with two symbols (SP) under channel-profile, mobility, and delay-spread mismatch for Exp1-Exp3.}
    \label{fig:ber_comparison}
\end{figure*}

\begin{figure}
    \centering
    \includegraphics[width=0.86\linewidth]{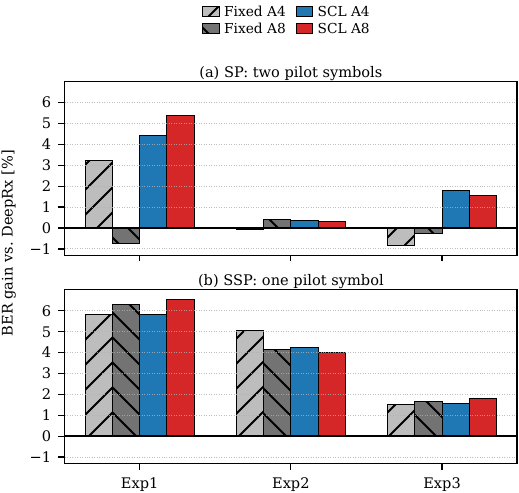}
    \caption{Anchor ablation for SP and SSP, measured as BER gain over DeepRx.}
    \label{fig:anch_ablation}
\end{figure}

Fig.~\ref{fig:ber_comparison} compares DeepRx, fixed-anchor \textsc{PARGAR}, channel-scaled \textsc{PARGAR}, LS-MMSE, and Genie-MMSE for the sparse pilot (SP) layout, where only 18 pilot REs are available. For readability, each \textsc{PARGAR} curve reports the post-evaluation lower-BER summary between $A_p=4$ and $A_p=8$ for the corresponding anchor rule. This summary is used because the two anchor counts often produce close BER curves; the full anchor-count comparison is reported separately in Fig.~\ref{fig:anch_ablation}. Therefore, the selected anchor count may differ across panels and between fixed-anchor and channel-scaled graphs.
In Fig.~\ref{fig:ber_comparison}(a), corresponding to Exp1, channel-scaled \textsc{PARGAR} gives the largest BER reduction relative to DeepRx and fixed-anchor \textsc{PARGAR}. This suggests that, under channel-profile mismatch, reduced pilot density makes pilot-anchor selection more sensitive to the effective time--frequency correlation structure, so the channel-scaled distance can be more informative than fixed geometric proximity.
Fig.~\ref{fig:ber_comparison}(b) shows the Exp2 speed-mismatch case. Here, all neural receivers remain close, and channel-scaled anchoring is nearly neutral relative to the fixed-anchor graph. Since the channel profile and delay spread are fixed in Exp2, the coherence-bandwidth term is unchanged and the coherence-time term mainly rescales time separation, providing limited additional discrimination for the same sparse DMRS layout.
Fig.~\ref{fig:ber_comparison}(c) presents the Exp3 delay-spread mismatch results. Channel-scaled anchoring provides lower BER than the fixed-anchor graph and yields a positive sparse-pilot gain, consistent with delay-spread variation changing frequency correlation and therefore the relevance of frequency-separated pilot anchors.
Overall, the SP results suggest that anchor count alone is insufficient: selected pilots must also be informative for the channel conditions. Channel-scaled anchoring is most useful when fixed geometric proximity is less representative of the channel-scaled time--frequency neighborhood, as in Exp1 and Exp3. LS-MMSE and Genie-MMSE serve as practical and perfect-CSI linear references, respectively.
\begin{figure*}[t]
    \centering
    \begin{subfigure}[t]{0.32\textwidth}
        \centering
        \includegraphics[width=\linewidth]{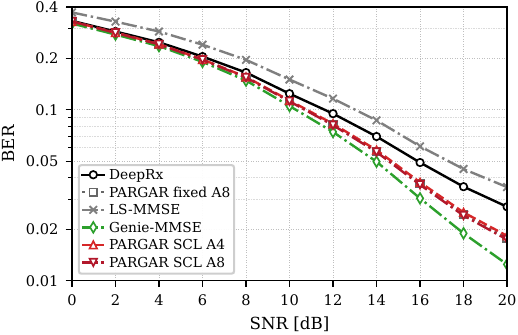}
        \caption{Exp1}
        \label{fig:exp1_p4_ber}
    \end{subfigure}
    \hfill
    \begin{subfigure}[t]{0.32\textwidth}
        \centering
        \includegraphics[width=\linewidth]{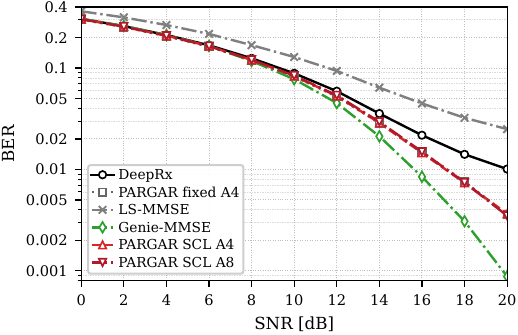}
        \caption{Exp2}
        \label{fig:exp2_p4_ber}
    \end{subfigure}
    \hfill
    \begin{subfigure}[t]{0.32\textwidth}
        \centering
        \includegraphics[width=\linewidth]{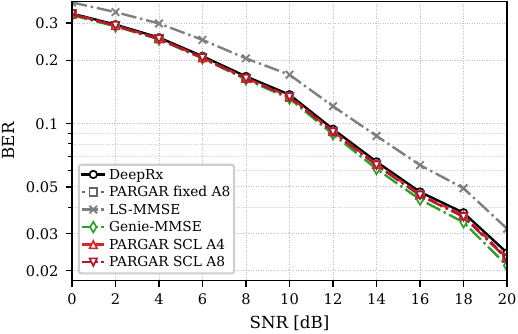}
        \caption{Exp3}
        \label{fig:exp3_p4_ber}
    \end{subfigure}
    \caption{Sparse-pilot with one symbol (SSP) under channel-profile, mobility, and delay-spread mismatch for Exp1-Exp3.}
    \label{fig:p4_ber_comparison}
\end{figure*}

Fig.~\ref{fig:p4_ber_comparison} reports the BER curves for the SSP pilot layout, which has the same pilot count as SP but concentrates the pilots on a single OFDM symbol. In Exp1, the best SSP \textsc{PARGAR} configuration is channel-scaled anchoring with $A_p=8$, reducing the average BER from 0.1490 for DeepRx to 0.1393, corresponding to a 6.54\% gain. In Exp2, the fixed-anchor $A_p=4$ configuration gives the lowest SSP \textsc{PARGAR} BER, reducing the average BER from 0.1181 to 0.1121, a 5.04\% gain. This is consistent with the earlier speed-mismatch results, in which channel scaling provides only limited additional discrimination because the channel profile and delay spread remain fixed. In Exp3, channel-scaled anchoring with $A_p=8$ gives a smaller but positive gain of 1.81\%, reducing the average BER from 0.1515 to 0.1488. Overall, the SSP results suggest that pilot-aware graph propagation remains useful when sparse pilots are concentrated in a single pilot-bearing OFDM symbol, with the largest observed benefit in the Exp1 channel-profile evaluation.

Fig.~\ref{fig:anch_ablation} reports the sparse-pilot anchor ablation as the percentage reduction in average uncoded BER relative to DeepRx. For a \textsc{PARGAR} configuration $m$, the plotted gain is computed as
\begin{equation}
    G_m =
    100\left(1-
    \frac{
    \frac{1}{|\mathcal{T}|}\sum_{(s,\xi)\in\mathcal{T}}
    \mathrm{BER}_{m}(s,\xi)}
    {
    \frac{1}{|\mathcal{T}|}\sum_{(s,\xi)\in\mathcal{T}}
    \mathrm{BER}_{\mathrm{DeepRx}}(s,\xi)}
    \right),
    \label{eq:relative_ber_gain}
\end{equation}
where $s$ denotes an evaluated SNR point, $\xi$ denotes a held-out test condition, and $\mathcal{T}$ is the set of evaluated SNR--test-condition pairs for the corresponding experiment. Positive values mean that \textsc{PARGAR} reduces BER compared with DeepRx, while negative values mean degradation. 
Fig.~\ref{fig:anch_ablation} separates the two sparse layouts: SP uses two pilot-bearing OFDM symbols, whereas SSP uses one pilot-bearing OFDM symbol with the same total pilot count. In Fig.~\ref{fig:anch_ablation}(a), the fixed-anchor $A_p=8$ degrades in Exp1, and both fixed-anchor settings remain below DeepRx in Exp3. Channel-scaled anchoring mitigates these cases, reaching a 5.36\% gain in Exp1-SP with $A_p=8$ and a 1.79\% gain in Exp3-SP with $A_p=4$. In Fig.~\ref{fig:anch_ablation}(b), all SSP anchor settings provide positive gains across the three experiments; the largest gain is 6.54\% in Exp1 with channel-scaled $A_p=8$, while Exp2 is strongest for fixed $A_p=4$ at 5.04\%. These results indicate that anchor count alone is insufficient: the physical relevance and temporal placement of selected pilot anchors also affect graph-based message passing. Future receivers could learn or adapt the anchor metric from pilot spacing, Doppler, delay spread, or validation feedback.
\begin{table}[t]
\centering
\caption{MMSE-feature ablation on Exp1 SP with fixed-anchor graphs.}
\label{tab:mmse_ablation}
\begin{tabular}{ccccc}
\toprule
    Anchor & DeepRx & PARGAR w/o MMSE & PARGAR & Gain \\ \midrule
    A4 & 0.1487 & 0.1512 & 0.1439 & 4.82\% \\
    A8 & 0.1487 & 0.1532 & 0.1498 & 2.24\% \\
    \bottomrule
\end{tabular}
\end{table}

Table~\ref{tab:mmse_ablation} isolates the MMSE-derived node features for the fixed-anchor graph in Exp1-SP. Without this feature, the graph-attention receiver has a higher BER than DeepRx. Adding the practical equalized-symbol feature reduces the \textsc{PARGAR} BER by 4.82\% for $A_p=4$ and 2.24\% for $A_p=8$, suggesting that the receiver benefits from both graph message passing and a model-based equalization prior. 
\vspace{-1 mm}
\section{Conclusion}
\label{sec:conclusion}
This paper presented \textsc{PARGAR}, a pilot-aware graph-attention receiver for OFDM bit detection under dense, two-symbol sparse, and single-symbol sparse pilot layouts. The main takeaway is that pilot-aware graph connectivity becomes more useful when pilot density is reduced, and pilot information must be propagated across larger time--frequency separations. Under dense pilots, \textsc{PARGAR} and DeepRx remain closely matched, suggesting that CNN-based and graph-based receivers can both exploit sufficiently frequent pilot observations in the considered setup. Under selected SP and SSP scenarios, channel-scaled pilot anchoring provides modest BER improvements by connecting data REs to pilot observations selected based on approximate coherence time and coherence bandwidth, whereas the SSP results show that the same pilot count can behave differently when pilot observations are concentrated in a single OFDM symbol. The study is limited to a SISO uncoded setting with coarse Doppler and delay-spread side information. Future work will consider MIMO, adaptive anchors, and inference efficiency.
\bibliographystyle{IEEEtran}
\bibliography{ref}

@article{velivckovic2017graph,
  title={Graph attention networks},
  author={Veli{\v{c}}kovi{\'c}, Petar and Cucurull, Guillem and Casanova, Arantxa and Romero, Adriana and Lio, Pietro and Bengio, Yoshua},
  journal={arXiv preprint arXiv:1710.10903},
  year={2017}
}

@article{honkala2021deeprx,
  title={{DeepRx}: Fully convolutional deep learning receiver},
  author={Honkala, Mikko and Korpi, Dani and Huttunen, Janne MJ},
  journal={IEEE Transactions on Wireless Communications},
  volume={20},
  number={6},
  pages={3925--3940},
  year={2021},
  publisher={IEEE}
}

@ARTICLE{11282490veritas,
  author={Soltani, Nasim and Löhning, Michael and Chowdhury, Kaushik},
  journal={IEEE Transactions on Mobile Computing}, 
  title={{VERITAS}: ${\underline{\text{V}}}$erifying the P${\underline{\text{er}}}$formance of A${\underline{\text{I}}}$-Native ${\underline{\text{T}}}$ransceiver ${\underline{\text{A}}}$ctions in Base-${\underline{\text{S}}}$tations}, 
  year={2026},
  volume={25},
  number={5},
  pages={6705-6719},
  doi={10.1109/TMC.2025.3641030}}

@inproceedings{cammerer2023neural,
  title={A neural receiver for {5G} {NR} multi-user {MIMO}},
  author={Cammerer, Sebastian and A{\"\i}t Aoudia, Fay{\c{c}}al and Hoydis, Jakob and Oeldemann, Andreas and Roessler, Andreas and Mayer, Timo and Keller, Alexander},
  booktitle={2023 IEEE Globecom Workshops (GC Wkshps)},
  pages={329--334},
  year={2023},
  organization={IEEE}
}

@inproceedings{wiesmayr2025standard,
  title={Design of a standard-compliant real-time neural receiver for {5G} {NR}},
  author={Wiesmayr, Reinhard and Cammerer, Sebastian and A{\"\i}t Aoudia, Fay{\c{c}}al and Hoydis, Jakob and Zakrzewski, Jakub and Keller, Alexander},
  booktitle={2025 IEEE International Conference on Machine Learning for Communication and Networking (ICMLCN)},
  pages={1--6},
  year={2025},
  organization={IEEE}
}

@article{berke2026site,
  title={Site-Specific Finetuning of Neural Receivers with Real-World 5G NR Measurements},
  author={Berke Baytekin, Nuri and Wiesmayr, Reinhard and Cammerer, Sebastian and Dick, Chris and Studer, Christoph},
  journal={arXiv e-prints},
  pages={arXiv--2603},
  year={2026}
}

@article{gupta2026neuromorphic,
  title={{NeuromorphicRx}: From Neural to Spiking Receiver},
  author={Gupta, Ankit and Dizdar, Onur and Chen, Yun and Kadan, Fehmi Emre and Sattarzadeh, Ata and Wang, Stephen},
  journal={IEEE Transactions on Wireless Communications},
  volume={25},
  pages={10161--10174},
  year={2026},
  publisher={IEEE}
}

@article{obeed2026continual,
  title={Learning During Detection: Continual Learning for Neural {OFDM} Receivers via {DMRS}},
  author={Obeed, Mohanad and Jian, Ming},
  journal={arXiv preprint arXiv:2602.20361},
  year={2026}
}

@inproceedings{norouzi2025graphnet,
  title={Lightweight graph neural networks for enhanced {5G} {NR} channel estimation},
  author={Norouzi, Sajedeh and Rahmani, Mostafa and Chu, Yi and Braun, Torsten and Chowdhury, Kaushik and Burr, Alister},
  booktitle={2025 IEEE 36th International Symposium on Personal, Indoor and Mobile Radio Communications (PIMRC)},
  pages={1--7},
  year={2025},
  organization={IEEE}
}

@INPROCEEDINGS{norouzi2025grinet,
  author={Norouzi, Sajedeh and Samikwa, Eric and Rahmani, Mostafa and Braun, Torsten and Burr, Alister},
  booktitle={2025 IEEE International Conference on Machine Learning for Communication and Networking (ICMLCN)}, 
  title={Decentralized Federated Learning for {GNN}-Based Channel Estimation With {DM-RS} in {O-RAN}}, 
  year={2025},
  volume={},
  number={},
  pages={1-7},
  doi={10.1109/ICMLCN64995.2025.11140565}}

@ARTICLE{rahmani2026survey,
  author={Rahmani Ghourtani, Mostafa and Norouzi, Sajedeh and Chen, Jinxuan and Ahmadi, Hamed and Braun, Torsten and Chowdhury, Kaushik and Burr, Alister},
  journal={IEEE Communications Surveys \& Tutorials}, 
  title={Hybrid {GNN}-Centric Architectures for {AI}-Native {6G} Wireless Networks: A Comprehensive Survey}, 
  year={2026},
  volume={28},
  number={},
  pages={5678-5712},
  doi={10.1109/COMST.2026.3681198}}

@ARTICLE{yang2020tracking,
  author={Yang, Yindi and Zhang, Shun and Gao, Feifei and Ma, Jianpeng and Dobre, Octavia A.},
  journal={IEEE Communications Letters}, 
  title={Graph Neural Network-Based Channel Tracking for Massive {MIMO} Networks}, 
  year={2020},
  volume={24},
  number={8},
  pages={1747-1751},
  doi={10.1109/LCOMM.2020.2990487}}

@article{hoydis2022sionna,
  title={Sionna: An open-source library for next-generation physical layer research},
  author={Hoydis, Jakob and Cammerer, Sebastian and Aoudia, Fay{\c{c}}al Ait and Vem, Avinash and Binder, Nikolaus and Marcus, Guillermo and Keller, Alexander},
  journal={arXiv preprint arXiv:2203.11854},
  year={2022}
}

@ARTICLE{11186253,
  author={Lin, Xingqin},
  journal={IEEE Communications Standards Magazine}, 
  title={A Tale of Two Mobile Generations: {5G}-Advanced and {6G} in {3GPP} Release 20}, 
  year={2025},
  volume={},
  number={},
  pages={1-9},
  doi={10.1109/MCOMSTD.2025.3613219}}

\end{document}